\documentclass[reprint, pra,
amsmath,amssymb,showpacs,superscriptaddress]{revtex4-1}
\pdfoutput=1

\usepackage[breaklinks=false]{hyperref}
\usepackage{graphicx}
\usepackage{braket}
\usepackage{verbatim}
\usepackage{float}
\usepackage[usenames,dvipsnames]{color}
\usepackage[amssymb]{SIunits}

\newcommand{\nm}{\nano\metre}

\newcommand{\THz}{\tera\hertz}

\newcommand{\MHz}{\mega\hertz}
\newcommand{\kHz}{\kilo\hertz}

\newcommand{\invcm}{\reciprocal{\centi\metre}}

\newcommand{\us}{\micro\second}

\newcommand{\uK}{\micro\kelvin}

\begin{document}
\title{Measurement of the binding energy of ultracold $^{87}$Rb$^{133}$Cs
molecules \\
using an offset-free optical frequency comb}

\author{Peter K. Molony}
\author{Avinash Kumar}
\author{Philip D. Gregory}
\affiliation{Joint Quantum Centre (JQC) Durham-Newcastle, Department of
Physics, Durham University, South Road, Durham DH1 3LE, United Kingdom}

\author{Russell Kliese}
\author{Thomas Puppe}
\affiliation{TOPTICA Photonics AG, Lochhamer Schlag 19, Gr\"{a}felfing 82166,
Germany}

\author{C. Ruth Le Sueur}
\affiliation{Joint Quantum Centre (JQC) Durham-Newcastle, Department of
Chemistry, Durham University, South Road, Durham, DH1 3LE, United Kingdom}

\author{Jesus Aldegunde}
\affiliation{Departamento de Quimica Fisica, Universidad de Salamanca, 37008
Salamanca, Spain}

\author{Jeremy M. Hutson}
\affiliation{Joint Quantum Centre (JQC) Durham-Newcastle, Department of
Chemistry, Durham University, South Road, Durham, DH1 3LE, United Kingdom}

\author{Simon L. Cornish}
\email{s.l.cornish@durham.ac.uk} \affiliation{Joint Quantum Centre (JQC)
Durham-Newcastle, Department of Physics, Durham University, South Road, Durham
DH1 3LE, United Kingdom}

\begin{abstract}
We report the binding energy of $^{87}$Rb$^{133}$Cs molecules in their
rovibrational ground state measured using an offset-free optical frequency comb
based on difference frequency generation technology. We create molecules in the
absolute ground state using stimulated Raman adiabatic passage (STIRAP) with a
transfer efficiency of 88\%. By measuring the absolute frequencies of our STIRAP
lasers, we find the energy-level difference from an initial weakly-bound
Feshbach state to the rovibrational ground state with a resolution of
$\unit{\sim5}{\kHz}$ over an energy-level difference of more than
$\unit{114}{\THz}$; this lets us discern the hyperfine splitting of the ground
state. Combined with theoretical models of the Feshbach state binding energies
and ground-state hyperfine structure, we determine a zero-field binding energy
of $h\times\unit{114\,268\,135\,237(5)(50)}{\kHz}$. To our knowledge, this is
the most accurate determination to date of the dissociation energy of a
molecule.

\end{abstract}

\maketitle

Quantum gases of polar molecules have received great attention in recent years.
Their long-range interactions and rich internal structure hold enormous
potential in the fields of quantum many-body
simulations~\cite{Santos:2000,Baranov:2012}, quantum
computation~\cite{DeMille:2002}, ultracold
chemistry~\cite{S.Ospelkaus_Science_2010,R.V.Krems_PCCP_2008} and precision
measurement of fundamental
constants~\cite{V.V.Flambaum_PRL_2007,T.A.Isaev_PRA_2010,J.J.Hudson_Nature_2011,
Demille2008}. It is only recently, however, that a limited selection of such
molecules (KRb, RbCs, NaK, NaRb) have been successfully trapped at ultracold
temperatures in their rovibrational ground
state~\cite{Ni_Science_2008,Takekoshi_PRL_2014, Molony_PRL_2014,
Park_PRL_2015,Guo2016}, making them available for experimental study. These
experiments all share a common technique for the production of molecules, in
which atoms are first associated to form weakly bound molecules by tuning a
magnetic field across a Feshbach resonance, and the molecules are then
transferred optically to the ground state using stimulated Raman adiabatic
passage (STIRAP)~\cite{Gaubatz_JCP_1990,K.Bergmann_PRA_1998}.

An accurate characterization of the internal structure of these molecules has
been challenging both theoretically and experimentally. The most precise
measurement so far of the binding energy of these molecules is for
KRb~\cite{Ni_Science_2008}, where a frequency comb was used to measure the
difference in laser frequency for the STIRAP transfer to a precision of
$\pm\unit{1}{\MHz}$ at a non-zero magnetic field. In $^{87}$Rb$^{133}$Cs, the
measurement precision has so far been approximately $\unit{20}{\MHz}$, limited
by the precision of wavemeters~\cite{M.Debatin_pccp_2011,Molony_PRL_2014}.

In this article, we present the most precise measurement of the binding energy
$D_0$, or dissociation energy, of the lowest rovibrational state of the
$^{87}$Rb$^{133}$Cs $X^1\Sigma^+$ ground-state potential to date. We begin with
a brief overview of the method we use to create samples of ultracold
ground-state $^{87}$Rb$^{133}$Cs molecules. We explain the working and stability
of our novel frequency comb based on difference frequency generation (DFG), and
how we use it to measure the $\unit{114}{\THz}$ frequency difference between the
STIRAP lasers. From this frequency difference we use theoretical models of the
molecular structure to calculate the binding energy of the $^{87}$Rb$^{133}$Cs
molecule at zero magnetic field.

\section{Creating ground-state molecules}\label{sec:CreatingMolecules}
\begin{figure}
	\centering
	\includegraphics[width=1.0\columnwidth]{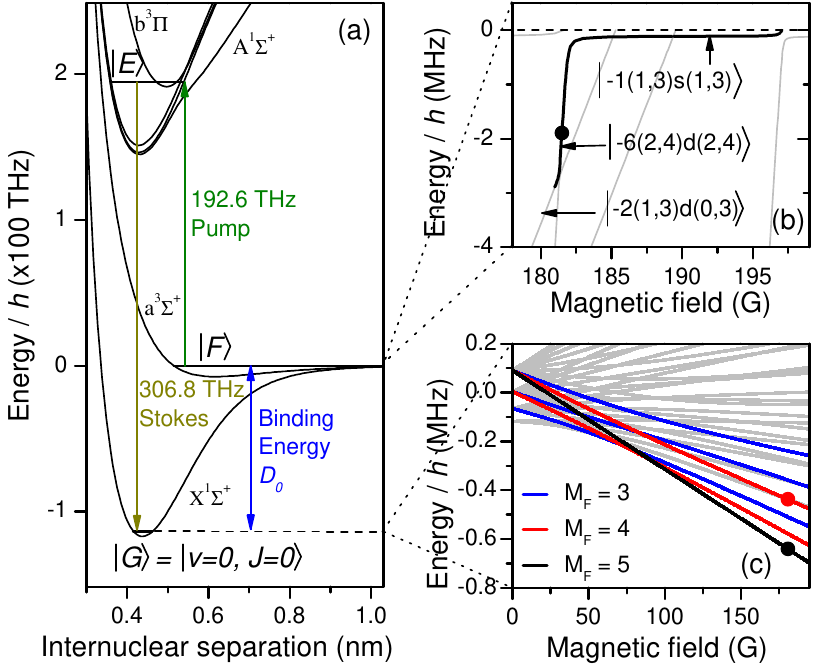}
\caption{(Color online) $^{87}$Rb$^{133}$Cs molecular states relevant to our
experiment. (a)~The position of the energy levels we use for STIRAP within the
molecular potentials. The initial Feshbach state, intermediate excited state and
ground state are labeled as $\ket{F}$, $\ket{E}$ and $\ket{G}$ respectively.
(b)~Molecular states close to dissociation. The dotted line is the Rb
$\ket{f=1,m_f=1}$ + Cs $\ket{3,3}$ threshold. The black line shows the path
followed by the molecules directly after magnetoassociation at the Feshbach
resonance at $\unit{197.10(3)}{G}$. (c) Zeeman splitting of the ground state
into 32 energy levels from total molecular nuclear spin $I''= 2,3,4$ and $5$.
Transitions to the highlighted states are allowed by selection rules. Dots
indicate the states we address with our laser system.}
	\label{Molecularpotentials}
\end{figure}

Details of our experimental setup may be found in our previous
publications~\cite{Harris_JPB_2008, Jenkin_EPJ_2011, Cho_EPJ_2011,
Mccarron_pra_2011,Koppinger_pra_2014}. Briefly, from a two-species
magneto-optical trap we load both species into a magnetic
trap~\cite{Harris_JPB_2008}. We use forced RF
evaporation~\cite{Jenkin_EPJ_2011}, followed by plain evaporation in a levitated
optical trap ($\lambda=\unit{1550}{\nm}$)~\cite{Cho_EPJ_2011}, to create a high
phase-space density mixture of ${\sim}3.0\times10^{5}$ atoms of each species at
a temperature of $\unit{\sim300}{\nano\kelvin}$~\cite{Mccarron_pra_2011}.
Molecules are produced from this atomic mixture by sweeping the magnetic field
across an interspecies Feshbach resonance at $\unit{197.10(3)}{G}$ at a rate of
$\unit{250}{G~\reciprocal{\second}}$~\cite{Koppinger_pra_2014}. After
magnetoassociation, molecules populate the near-threshold $\ket{-1(1,3)s(1,3)}$
spin-stretched bound state of the potential $a^{3}\Sigma^{+}$ as shown in
figure~\ref{Molecularpotentials}(b). Here, states are labeled as
$\ket{n(f_{\rm{Rb}}, f_{\rm{Cs}})L(m_{f_{\rm{Rb}}},m_{f_{\rm{Cs}}})}$, where $n$
is the vibrational quantum number counted downward from the dissociation
threshold for the particular hyperfine $(f_{\rm{Rb}}, f_{\rm{Cs}})$ manifold,
and $L$ is the standard letter designation for the molecular rotational angular
momentum quantum number~\cite{Takekoshi_pra_2012}. We transfer our molecules to
the weakly bound $\ket{-2(1,3)d(0,3)}$ state by reducing the magnetic field to
$\unit{\sim180.5}{G}$, at which point the atoms and molecules are separated
using the Stern-Gerlach effect (at a field gradient of
$\unit{44}{G~\reciprocal{\centi\metre}}$), taking advantage of their different
magnetic moments when the molecules are in this state. We then reduce the
magnetic field gradient and ramp up the dipole trap to create a pure optical
trap. Finally, the magnetic field is ramped to $\unit{\sim181.5}{G}$ to transfer
the molecules into a state which is suitable for transfer to the rovibrational
ground state. This results in ${\sim}2500$ molecules in the
$\ket{-6(2,4)d(2,4)}$ state at a temperature of $\unit{1.5}{\uK}$.
The weakly bound molecules are transferred to the rovibrational ground state
optically using STIRAP. We couple both the initial near-dissociation state and
the ground state to a common excited state. This excited state is chosen to be
the $\ket{\Omega'=1, v'=29, J'=1}$ state, from the coupled
$A^{1}\Sigma^{+}+b^{3}\Pi$ potential, because it has strong couplings to both
the Feshbach and ground states~\cite{Takekoshi_PRL_2014}. The pump and Stokes
lasers are shown schematically in figure~\ref{Molecularpotentials}(a), and have
frequencies of $\unit{192.6}{\THz}$ ($\unit{1557}{\nm}$) and
$\unit{306.8}{\THz}$ ($\unit{977}{\nm}$) respectively. For coherent transfer, we
narrow the linewidth of both pump and Stokes lasers to $<\unit{1}{\kHz}$ by
frequency stabilisation to a fixed-length high-finesse optical cavity
constructed from ultra-low-expansion (ULE) glass by ATFilms. Continuous tuning
is given by a pair of fibre-coupled electro-optic modulators. Further details of
the laser system can be found in~\cite{Gregory_NJP_2015}.
\begin{figure}
	\centering
	\includegraphics[width=1.0\linewidth]{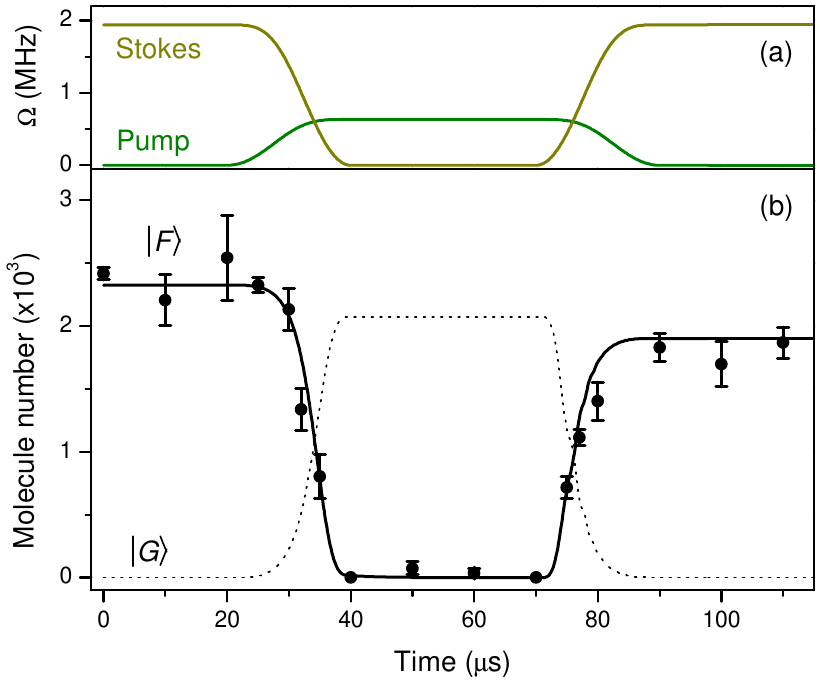}
\caption{(Color online) STIRAP transfer to the molecular ground state and back.
(a) Rabi frequency profile used for STIRAP transfer. (b) Experimentally measured
population of the Feshbach state $\ket{F}$ throughout the transfer process. The
sequence keeps the molecules in the ground state for $\unit{30}{\us}$ before
transferring them back to the initial state for dissociation and absorption
imaging. We show a numerical model of the Feshbach and ground state populations
based on the Lindblad master equation for an open three-level system, including
the effects of laser linewidth. The one-way transfer efficiency is 88\%. The
optical trap is switched off throughout the sequence. }
	\label{PulseSeq}
\end{figure}

We transfer the molecules to the ground state and back as shown in
figure~\ref{PulseSeq}. This figure shows a model of the Lindblad master equation
for an open three-level system, using our measured peak Rabi frequencies of
$\unit{0.6}{\MHz}$ and $\unit{1.9}{\MHz}$ for the pump and Stokes transitions
respectively. The details of this model will be presented in a separate
publication. As the dipole trapping wavelength is close to the pump transition,
it induces an AC Stark shift of $\unit{\sim0.5}{\MHz}$. This shift varies across
the cloud because of the finite size of the molecular cloud and trapping beams,
reducing the efficiency of the transfer. To avoid this we switch our dipole trap
off for $\unit{200}{\us}$ during the STIRAP transfer to and from the ground
state. This improves our one-way transfer efficiency from
50\%~\cite{Molony_PRL_2014} to 88\%, creating a sample of over 2000 molecules in
the rovibrational ground state.

\section{Laser frequency measurement}\label{sec:FrequencyComb}
We determine the binding energy with precision measurements of the pump and
Stokes transition frequencies using a GPS-referenced frequency comb. Our
frequency comb is the first of its kind, based on difference frequency
generation technology developed by TOPTICA Photonics AG~\cite{Telle:2005}. In
this comb, the amplified output of an Er:fiber oscillator is compressed using a
silicon prism compressor and then spectrally broadened using a highly nonlinear
photonic crystal fiber to make a supercontinuum spanning more than an optical
octave. The comb teeth in the spectrum are given by
$f=Nf_{\textrm{rep}}+f_{\textrm{CEO}}$. Two extreme parts of this supercontinuum
are spatially and temporally overlapped in a nonlinear difference frequency
generation (DFG) crystal. This cancels the carrier-envelope offset frequency
($f_{\textrm{CEO}}$) to produce an offset-free frequency comb spectrum at
$\unit{1550}{\nm}$ with a bandwidth of $\unit{\sim100}{\nm}$. Each comb tooth
$N$ then has a frequency $f=Nf_{\textrm{rep}}$. This output is then extended to
different wavelength ranges by nonlinear frequency shifting and frequency
doubling. This method to cancel $f_{\textrm{CEO}}$ has the advantage of
requiring no servo-loop feedback system, compared to the conventional $f-2f$
approach where the high-frequency noise components of $f_{\textrm{CEO}}$ cannot
be canceled~\cite{Fuji2004}. The characterization of the phase noise of
different comb teeth confirms the elastic tape
model~\cite{E.Benkler_OpticsExpress_2005} with a fixed point at zero
frequency~\cite{Puppe2016}.

\begin{figure}
	\centering
	\includegraphics[width=1.\columnwidth] {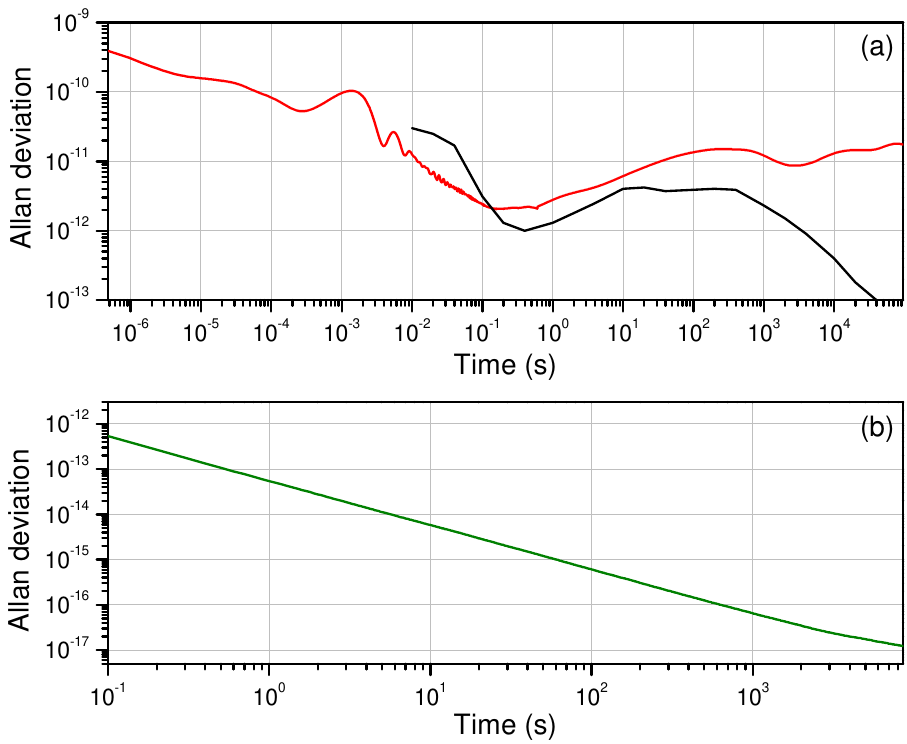}
\caption{(Color online) Measurements of the stability of the frequency comb.
(a)~Allan deviation (AD) of a beat between the frequency comb and a laser
stabilized to the Rb 5S$_{1/2}(f$=2)$\rightarrow$5P$_{3/2}(f'$=3) transition
(red), and the AD of the GPS-referenced $\unit{10}{\MHz}$ oscillator (black) to
which the comb is locked. (b)~AD of the beat signal between two identical DFG
combs locked to a common reference. The beat was recorded at a wavelength of
$\unit{1556}{\nm}$ via a transfer oscillator~\cite{Puppe2016}.}
	\label{FrequencyComb}
\end{figure}
The frequency comb is seeded by a mode-locked Er:fibre laser with an
$\unit{80}{\MHz}$ repetition rate, whose 10th harmonic is locked to an
$\unit{800}{\MHz}$ ultra-low-noise oven-controlled RF oscillator, which in turn
is locked to a $\unit{10}{\MHz}$ GPS reference (Jackson Labs Fury). We have
measured the absolute stability of the comb locked to the GPS reference by
recording a beat note between a comb tooth and a laser stabilized to the Rb
5S$_{1/2}(f$=2)$\rightarrow$5P$_{3/2}(f'$=3) line. Figure \ref{FrequencyComb}(a)
shows the Allan deviation (AD) of the beat signal, compared to the AD of the GPS
referenced oscillator to which the comb is locked. The AD of the beat follows a
similar trend to the reference signal but deviates at longer time scales. This
deviation is due to the drift in the lock-signal offset of the laser locked to
the Rb spectroscopy line and is commonly observed over such time scales. These
results show that measuring uncertainties down to $10^{-11}$ is practical with
our comb system.

To quantify the lock noise of the comb, we measure the AD of a beat signal
between two combs locked to a common RF reference. We observe an overall AD
lower than the reference signal with no similarity to the AD of the reference
signal (figure~\ref{FrequencyComb}(b)). This indicates that the AD of the
reference RF is completely canceled in this measurement and the AD related to
lock noise is much smaller than that. Therefore we can consider the AD of the
GPS signal at time scales greater than our experimental cycle to calculate the
resulting deviation on the repetition rate.

The frequency difference between the two STIRAP lasers is measured with comb
teeth separated by $\delta
N=\unit{(306.8-192.6)}{\THz}/\unit{80}{\MHz}\sim10^{6}$, so the uncertainty in
the GPS clock frequency must be less than $\unit{10}{\milli\hertz}$ if we are to
maintain an uncertainty in our measured laser frequency of less than
$\unit{10}{\kHz}$. The AD over time scales shorter than the experimental cycle
will add to the statistical error of the molecular round-trip signal. However,
the AD over longer time scales will lead to a systematic offset in our
measurements. From the specifications of the GPS reference we calculate that,
over the course of one measurement, the AD leads to a systematic uncertainty of
$\unit{\pm23}{\hertz}$ on the frequency difference between the two lasers. This
is negligible compared to the other sources of uncertainty described later.
\begin{figure}
	\centering
\includegraphics[trim={1cm 1.5cm 1cm
1.0cm},clip=true,width=0.98\columnwidth]{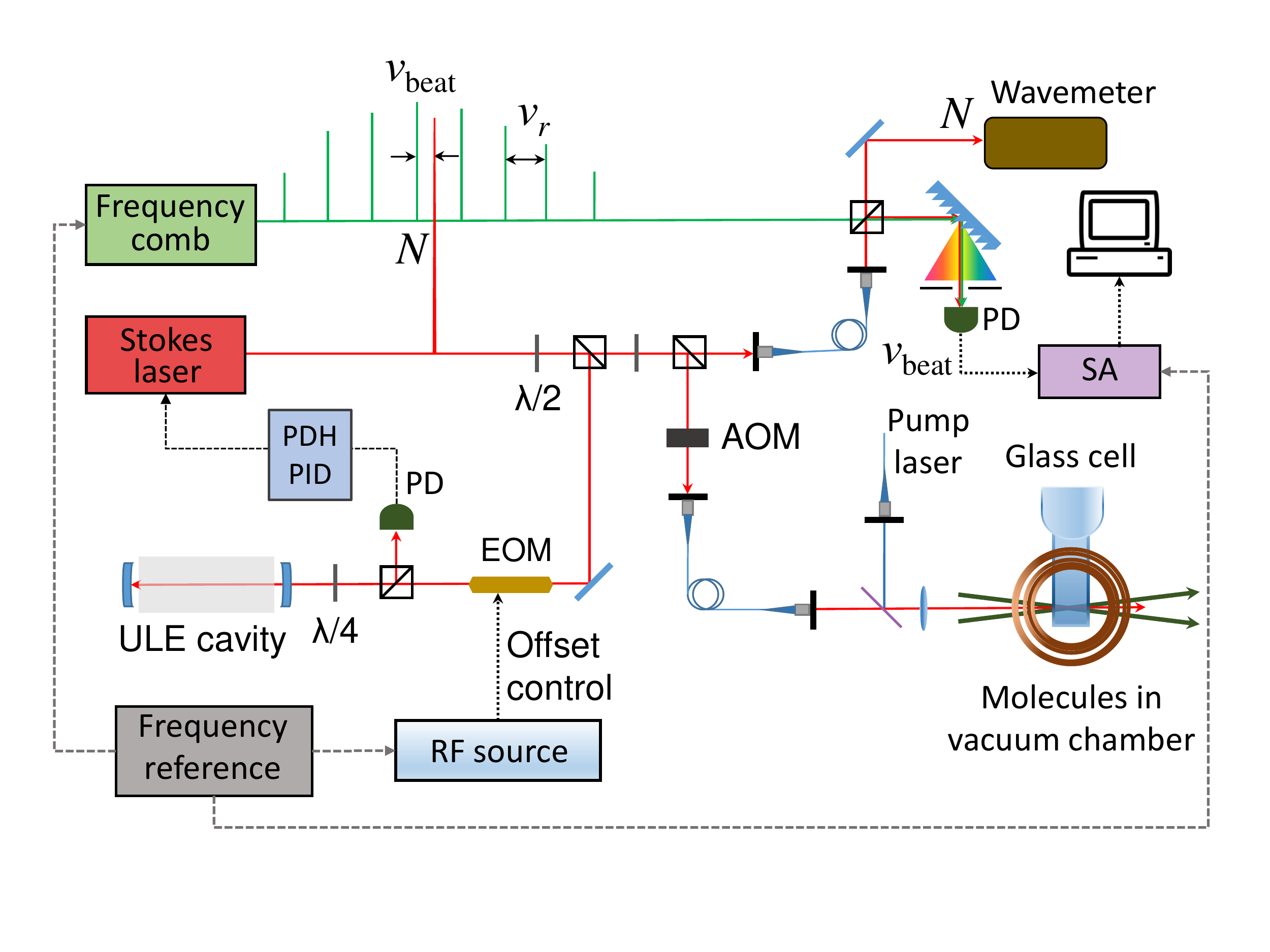}
\caption{(Color online) Schematic diagram of the experiment to carry out
spectroscopy while recording the beat note ($\nu_{\text{beat}}$) of the STIRAP
lasers with the optical frequency comb. The beat signal between each STIRAP
laser and the nearest comb line ($N$) is detected on a photodiode which is
connected to a spectrum analyzer (SA). Both STIRAP lasers are frequency
stabilized to a common ULE cavity using the Pound-Drever-Hall
method~\cite{Black_AJP_2001}. Continuous tuning of each laser is provided by
varying the RF driving frequency of a broadband fibre-coupled electro-optic
modulator (EOM). The light reaching the molecules is offset by $\unit{80}{\MHz}$
from that sent to the frequency comb by an acousto-optic modulator (AOM) which
modulates the intensity of the light. Further details of the frequency
stabilization and tuning of the STIRAP lasers can be found in
\citet{Gregory_NJP_2015}. The frequency comb, spectrum analyzers and EOM driver
are all referenced to the same $\unit{10}{\MHz}$ GPS disciplined oscillator. The
figure shows the setup for the Stokes laser; the setup for the pump laser is
identical.}
	\label{Schematic}
\end{figure}

The absolute frequency of the lasers is measured by beating light from each of
the STIRAP lasers with the nearest tooth of the optical frequency comb. A
schematic diagram of the optical setup used to measure the beat note and the
comb tooth number is shown in figure~\ref{Schematic}. The beat note is recorded
on a spectrum analyzer (Agilent N9320B for the Stokes, Agilent N1996 for the
pump), which is referenced to the same $\unit{10}{\MHz}$ GPS clock as the comb.
The frequency of the beat note is averaged and recorded over each three-second
interval. We identify the nearest comb tooth ($N$) using a wavemeter with an
absolute accuracy of $\unit{30}{\MHz}$ (High Finesse WS-U), which we calibrate
with lasers locked to well-known spectral lines in Rb, Cs and Sr.

The light reaching the molecules is offset from that sent to the frequency comb
by a pair of acousto-optic modulators (AOMs), at $\unit{+80}{\MHz}$ and
$\unit{-80}{\MHz}$ for the pump and Stokes respectively. These provide the
analog intensity ramps for STIRAP, and are driven by ISOMET 532B fixed-frequency
driver/amplifiers. We measure the accuracy of the absolute frequency of these
drivers on a spectrum analyser (Agilent N1996 referenced to the
$\unit{10}{\MHz}$ GPS clock) and find a constant offset of
$\unit{-705.0(3)}{\hertz}$ from the nominal $\unit{80}{\MHz}$. The statistical
uncertainty on this offset is negligible.

\section{Energy difference measurement}\label{sec:Method}
Maximum STIRAP transfer efficiency is achieved when the laser frequencies meet
the two-photon resonance condition, while any common detuning of both lasers has
relatively little effect on the efficiency~\cite{K.Bergmann_PRA_1998,
Gregory_NJP_2015}. By scanning their frequency difference and observing where we
get maximum transfer efficiency, we determine the energy difference between the
initial state $\ket{F}$ and final state $\ket{G}$.

\begin{figure}
	\centering
	\includegraphics[width=1.0\columnwidth]{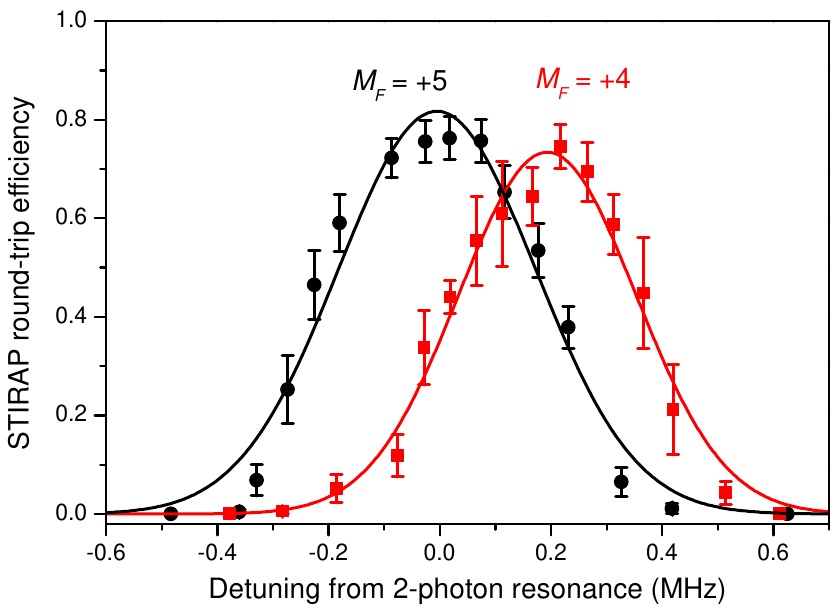}
\caption{(Color online) STIRAP transfer to different hyperfine sub-levels. The
STIRAP round-trip transfer efficiency changes with the frequency difference of
the pump and Stokes lasers for horizontal (black circles) and vertical (red
squares) Stokes polarizations at a magnetic field of $\unit{\sim181.5}{G}$.
Black circles show the transfer to the ${M_{F}=+5}$ state, while red squares
show the ${M_{F}=+4}$ state. Gaussian fits give a separation between the states
of $\unit{0.194(10)}{\MHz}$.}
	\label{fig_VH1_VV1}
\end{figure}

To measure the energy difference, we fix the frequency of the pump laser on
resonance with the Feshbach and intermediate states. We then vary the frequency
of the Stokes laser and measure the efficiency of the STIRAP transfer. The beat
notes of both lasers with the optical frequency comb are measured throughout.
For each data point we subtract the pump and Stokes absolute frequencies
measured with the comb, and add the shifts from the AOMs, to get an absolute
frequency difference. This gives us a peak as a function of Stokes frequency
which we fit to determine the energy difference between the initial and final
states, as shown in figure~\ref{fig_VH1_VV1}. The optimal Stokes frequency is
determined over $\sim4$ hours.

The precision with which we can locate the two-photon resonance is limited by
the shot-to-shot noise in the number of molecules which we produce. This noise
results in the vertical error bars seen in figure~\ref{fig_VH1_VV1}. The
uncertainties in the detuning (the horizontal error bars) are too small to be
seen. A Gaussian fit gives an uncertainty on the center of the spectroscopic
feature of around $\unit{\pm8}{\kHz}$. The magnetic field is measured before and
after each complete measurement using the microwave transition frequency between
the $\ket{f=3, m_{f}=+3}$ and $\ket{4,+4}$ states in atomic Cs.

We found the same frequency difference between the pump and Stokes transition,
within our experimental uncertainty, when using $\ket{\Omega'=0, v'=35, J'=1}$
as an alternative intermediate state. This measurement was carried out using
two-photon spectroscopy (where both the pump and Stokes light are pulsed on
simultaneously) as the coupling strengths are not high enough for efficient
STIRAP transfer. The experimental procedure for the two-photon spectroscopy of
the ground state has been discussed previously by \citet{Molony_PRL_2014} and
\citet{Gregory_NJP_2015}. This method, and the different transition strengths
and linewidths, results in a much wider spectroscopic signal, leading to much
larger uncertainties on the two-photon resonance.

\section{Binding Energy Calculation}\label{sec:BECalculation}
We will now combine the measured energy difference and magnetic field with
theoretical models, to determine the energy difference between the
degeneracy-weighted centres of the atomic and molecular hyperfine manifolds. We
must correct for several shifts which are included in our measurement: the
atomic hyperfine splittings, the Zeeman shifts of the $\ket{1,1}$ and
$\ket{3,3}$ atomic states, the binding energy of the Feshbach molecule relative
to these atomic states, and the molecular ground-state hyperfine splitting and
Zeeman shift. The effects of all of these shifts are summarised in
table~\ref{tab:AllErrors}. We will discuss each of these below.
\begin{table}
	\begin{ruledtabular}
		\begin{tabular}{lrr}
			Source &Correction (MHz)\phantom{\,000} &Error (MHz)\\
			\hline
$\nu_{\text{Stokes}}-\nu_{\text{pump}}$ & 114\,258\,363.067\phantom{\,000} &
0.006 \\
			\hline
			Feshbach binding energy&1.838\phantom{\,000} &\\
			Rb Zeeman&194.084\phantom{\,000}&\\
			Cs Zeeman&134.353\phantom{\,000}&\\
			RbCs Zeeman& $-$0.734\phantom{\,000} &\\
			Total Zeeman&$\phantom{-330.609}\phantom{\,000}$&0.013\\
			\hline
			Cs hyperfine&$\frac{9}{16}\times9\,192.631\,770$&$\equiv 0$\\
			Rb hyperfine&$\frac{5}{8}\times6\,834.682\,611$&$<10^{-10}$\\
			RbCs hyperfine ($I$=5)&0.091\phantom{\,000}&\\
			\hline
			Binding energy& 114\,268\,135.230\phantom{\,000} &0.014 \\
		\end{tabular}
	\end{ruledtabular}
\caption{\label{tab:AllErrors} All the corrections, and their respective
experimental errors, which must be added to our measurement of the energy
difference $\nu_{\text{Stokes}}-\nu_{\text{pump}}$ to give the energy difference
between the degeneracy-weighted hyperfine centroids of the free atoms and the
RbCs rovibrational ground state, \emph{i.e.} the binding energy. The uncertainty
in the Zeeman shift is from the uncertainty in the measured magnetic field.
Additional systematic uncertainties apply as explained in the text. The values
shown are from the second measurement in figure~\ref{FinalValues} at a magnetic
field $\unit{181.538(6)}{G}$ driving a transition to the $M_{F}=5$ hyperfine
ground state. All values are in MHz.}
\end{table}

The Cs ground-state hyperfine splitting at zero field comes directly from the
definition of the second, while the Rb splitting has been measured to
$<\unit{100}{\micro\hertz}$~\cite{Bize1999}. These are weighted by the
degeneracies of the atomic hyperfine states to give the distance to the
$5^2S_{1/2}+6^2S_{1/2}$ center. The atomic Zeeman splittings are calculated from
the standard atomic Hamiltonian. The electron spin, electron orbital and nuclear
\emph{g}-factors are the CODATA recommended values~\cite{CODATA2014}. We assume
the theoretical errors on these models are negligible.
\begin{figure}
	\centering
	\includegraphics[width=1.0\columnwidth] {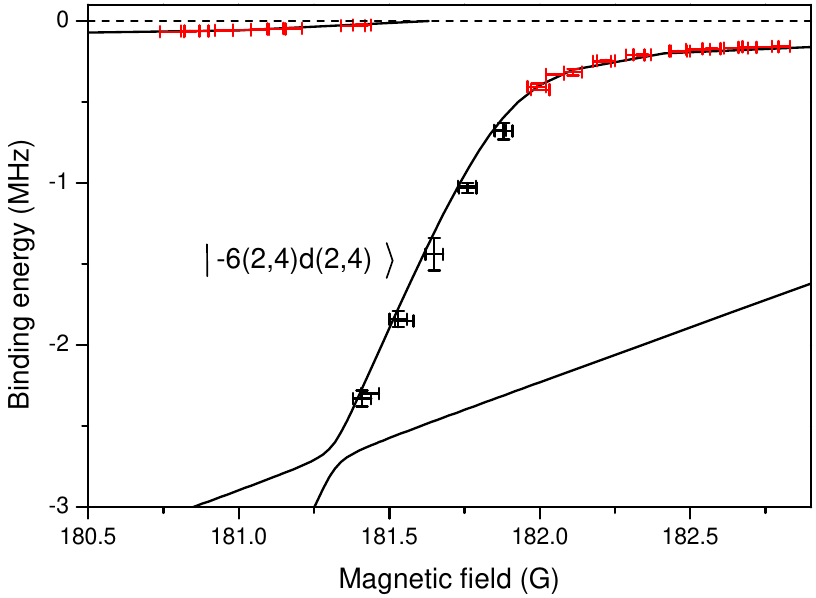}
\caption{(Color online) The calculated positions of the highest-lying bound
states for $^{87}$Rb$^{133}$Cs (solid black lines) together with the positions
measured by free-bound magnetic-field modulation spectroscopy. The measurements
included in the analysis of the required shifts of the binding energy (see main
text) are colored black, and the other data points in the set are colored red.
There are 9 points included in the fit; some of them nearly overlap. Data taken
from \citet{Takekoshi_pra_2012}.}
	\label{FeshbachBindingEnergy}
\end{figure}

We estimate the binding energy of the Feshbach state with respect to the
$\ket{1,1} + \ket{3,3}$ threshold by combining the measurements and the
coupled-channel model of reference~\cite{Takekoshi_pra_2012}, as shown in
figure~\ref{FeshbachBindingEnergy}. There are 9 experimental points for the
$|-6(2,4)d(2,4)\rangle$ state between 181.4~G and 181.9~G , and the
coupled-channel model systematically underestimates the binding energies by
$\unit{0.09(4)}{\MHz}$. For the present work, we recalculate the binding
energies from the coupled-channel model as a function of $B$ and increase the
resulting binding energies by this amount. We include the $\unit{40}{\kHz}$
uncertainty as a theoretical contribution to the final value for the
ground-state binding energy.

The $J=0$ rovibrational ground state has 4 hyperfine levels with nuclear spins
$I=2,3,4,5$. In the presence of a magnetic field, these are split into 32
hyperfine and Zeeman states originating from the nuclear spin coupling to the
magnetic field. These energy levels were calculated using the molecular
Hamiltonian and parameters in reference~\cite{Aldegunde_PRA_2008} and are
plotted in figure~\ref{Molecularpotentials}(c). We subtract both the hyperfine
and the Zeeman shifts to give the binding energy of the ground-state hyperfine
centroid, i.e. the zero of the energy axis in
figure~\ref{Molecularpotentials}(c).

There are also theoretical uncertainties associated with the model of the
ground-state hyperfine structure. The hyperfine splitting of the $I=2,3,4,5$
states is determined almost entirely by the scalar nuclear spin-spin coupling
constant $c_4$, which was calculated using density-functional theory (DFT) by
\citet{Aldegunde_PRA_2008}. We estimate that the uncertainty on $c_4$ is
$\pm30\%$, giving an uncertainty of $\unit{\pm27}{\kHz}$ on the position of the
$I=5$ state relative to the degeneracy-weighted hyperfine centroid. The Zeeman
shift is determined by the nuclear shielding constants, also from DFT
\cite{Aldegunde_PRA_2008}, but we estimate that the uncertainties in these
shieldings cause an uncertainty of only $\unit{\pm1}{\kHz}$. We combine these
ground-state uncertainties with the theoretical uncertainty on the model of the
Feshbach binding energy to give a total theoretical error of $\unit{50}{\kHz}$.
This is included as a separate ``theoretical'' uncertainty in the final value of
the ground-state binding energy.

We selectively address different hyperfine sublevels of the rovibrational ground
state by changing the polarization of the Stokes laser~\cite{Takekoshi_PRL_2014}
while keeping the pump laser polarization fixed parallel to the quantization
axis. The weakly bound state from which we begin our STIRAP transfer has a total
angular momentum projection quantum number~$M_F=+4$. In the case of Stokes
polarization parallel to the quantization axis, we drive $\pi$ transitions and
address a ground state where the $M_F$ value is unchanged. If, on the other
hand, the Stokes polarization is perpendicular to the quantization axis, we
drive $\sigma^\pm$ transitions and address ground states with either $M_F=+3$ or
$M_F=+5$.

In figure~\ref{fig_VH1_VV1}, we see the effect of scanning the Stokes laser
frequency on the efficiency of STIRAP transfer for both parallel and
perpendicular polarizations. The coupling strengths to the hyperfine ground
states are such that we have sufficient laser power to populate only two of the
available hyperfine states, which are separated in energy by
$\unit{0.194(10)}{\MHz}$. The measured energy difference, in combination with
knowledge of the states accessible with different Stokes polarizations, allows
us to identify the two states as indicated in
figure~\ref{Molecularpotentials}(c), agreeing with previous
results~\cite{Takekoshi_PRL_2014}. Both of these Zeeman states correlate with
the $I=5$ hyperfine state. Because of mixing between the $I=4$ and $I=5$ states
in a magnetic field, the measured splitting of $\unit{0.194(10)}{\MHz}$ has some
dependence on the spin-spin coupling constant $c_4$. It corresponds to a value
$c_4=\unit{0.023(7)}{\kHz}$, which agrees within its error bars with the value
of $\unit{0.01734}{\kHz}$ from DFT calculations \cite{Aldegunde_PRA_2008} and is
also consistent with our attribution of an uncertainty of 30\% to the latter
value. We note that at a field of $\unit{\sim181.5}{G}$ the $M_F=+5$ state is
the lowest-energy sublevel, as shown in figure~\ref{Molecularpotentials}(c).

We must also consider the effect of the uncertainty in the magnetic field. We
have considered the atomic and molecular Zeeman shifts separately above, but
with the uncertainty in the field they must be considered together. We multiply
the uncertainty in the measured field by the difference in magnetic moment
between the Feshbach and ground states to give the associated uncertainty in the
binding energy. This is shown in table~\ref{tab:AllErrors}, and is added to the
uncertainty from the frequency difference measurement above to give the total
statistical uncertainty on the binding energy.

\section{Measurement campaign}\label{sec:Campaign}
\begin{figure}
	\centering
	\includegraphics[width=1.0\columnwidth] {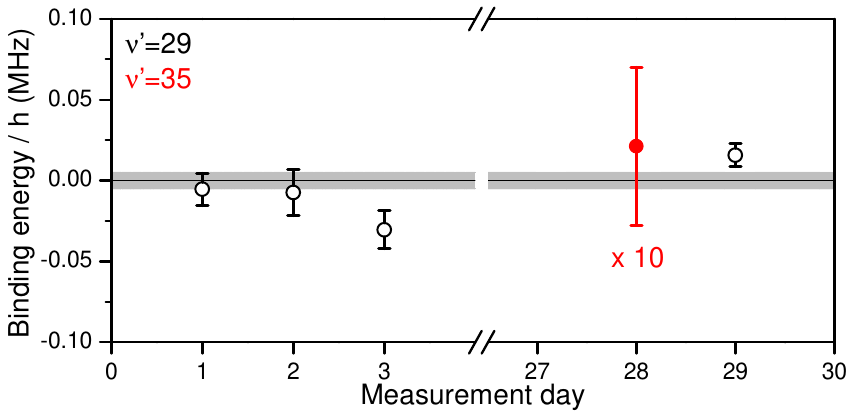}
\caption{(Color online) Binding energy of the $^{87}$Rb$^{133}$Cs molecule
measured on different days, with experimental error bars. The vertical scale is
offset by the mean value of $\unit{114\,268\,135.237}{\MHz}$. The gray shaded
region represents the $\unit{5}{\kHz}$ experimental error on the mean. Black
(red) data points show the binding energy calculated from two-photon
spectroscopy via $\ket{\Omega'=1, v'=29, J'=1}$ ($\ket{\Omega'=0, v'=35, J'=1}$)
as the intermediate state. The $\ket{\Omega'=0, v'=35, J'=1}$ measurement and
error bar have been divided by 10 for clarity. The larger experimental errors in
two-photon spectroscopy via the $\ket{\Omega'=0, v'=35, J'=1}$ state are due to
the poor signal-to-noise ratio of the molecular spectroscopy signal.}
	\label{FinalValues}
\end{figure}

We have repeated the measurement outlined in the section~\ref{sec:BECalculation}
five times on different days, and observed similar results for the energy
difference each time, within experimental errors. In this section, we combine
these measurements to give a value for the binding energy $D_0$. All five
measurements are summarised in figure~\ref{FinalValues}, and the precise values
for each measurement are shown in table~\ref*{tab:Freq}.

The measurement shown in red in figure~\ref{FinalValues} uses the
$\ket{\Omega'=0, v'=35, J'=1}$ intermediate state. The polarisations are such
that we expect to address the $M_F=3,5$ states, but the large spectroscopic
linewidth means this measurement does not resolve the ground-state hyperfine
structure. The main purpose of this measurement is to confirm that we have
identified the frequency comb tooth correctly. The other four measurements use
the $\ket{\Omega'=1, v'=29, J'=1}$ state in the coupled
$A^{1}\Sigma^{+}+b^{3}\Pi$ potential. Of these, three are measured with the
$M_F=5$ ground-state hyperfine level, and one uses the $M_F=4$ state.

Following the procedure in the previous sections, we calculate values for
binding energies for each measurement. Taking a weighted mean we get a final
value for the binding energy of $^{87}$Rb$^{133}$Cs of
\begin{align*}
D_0&=h\times\unit{114\,268\,135\,237(5)(50)}{\kHz}\\
&=hc\times\unit{3811.574\,714\,03(16)(200)}{\invcm}.
\end{align*}
The first uncertainty arises from the statistical experimental error and the
second one arises from the theoretical uncertainties in the coupled-channel
model and the ground-state hyperfine splitting. The uncertainty in the final
value of the binding energy is dominated by the theoretical uncertainties. Our
experimental frequency measurements are more accurate by one order of magnitude.
\begin{table*}
	\begin{ruledtabular}
		\begin{tabular}{lllll}
Polarization & $M_F$ & $\nu_{\text{Stokes}}-\nu_{\text{pump}}$ (MHz) &
$B$~(G)&$D_0/h$ (MHz)\\
			\hline
$V_{\text{P}}, V_{\text{S}}$ & 4 & 114\,258\,362.874(8) & 181.542(3) &
114\,268\,135.232(10)\\
$V_{\text{P}}, H_{\text{S}}$ & 5 & 114\,258\,363.067(6) & 181.538(6) &
114\,268\,135.230(14)\\
$V_{\text{P}}, H_{\text{S}}$ & 5 & 114\,258\,363.075(8) & 181.552(4) &
114\,268\,135.207(12)\\
$V_{\text{P}}, H_{\text{S}}$* & 5 & 114\,258\,363.2(5) & 181.510(3) &
114\,268\,135.4(5)\\
$V_{\text{P}}, H_{\text{S}}$ & 5 & 114\,258\,363.048(5) & 181.519(2) &
114\,268\,135.253(7)\\
		\end{tabular}
	\end{ruledtabular}
\caption{\label{tab:Freq} Summary of each independent measurement of the binding
energy in the ground state. Both the magnetic field and the polarization of the
pump light are vertical ($V_{\text{P}}$). The Stokes light may be either
vertical ($V_{\text{S}}$) or horizontal ($H_{\text{S}}$) to access ground-state
hyperfine levels with either $M_F=4$ or $M_F=5$. For each measurement we show
the absolute frequency difference measured for each laser
($\nu_{\text{Stokes}}-\nu_{\text{pump}}$), the magnetic field during that
measurement ($B$), and the binding energy of the ground state at zero field
($D_0$). An additional $\unit{0.05}{\MHz}$ theoretical uncertainty applies to
the binding energies, as explained in the text. An asterisk* indicates a
measurement using two-photon spectroscopy via the intermediate $\ket{\Omega'=0,
v'=35, J'=1}$ state. All other measurements rely on optimization of the
round-trip STIRAP efficiency via the intermediate $\ket{\Omega'=1, v'=29, J'=1}$
state.}
\end{table*}

This value is a 500-fold improvement in accuracy over previous measurements
averaging
$\unit{3811.5759(8)}{\invcm}$~\cite{M.Debatin_pccp_2011,Molony_PRL_2014}. The
most precise determinations of a molecular binding energy we know of are
precisions of $\Delta E/E\sim10^{-8}$. These are in $^{40}$K$^{87}$Rb, which is
measured with $8\times10^{-9}$ precision at a finite magnetic
field~\cite{Ni_Science_2008}, and $\textrm{H}_2$ with $1\times10^{-8}$
precision~\cite{Liu2009}. Our fractional uncertainty is $4\times10^{-10}$, and
improved models and measurements of the Feshbach and ground-state structure
could reduce this as far as $5\times10^{-11}$.

\section{Conclusions}\label{sec:Conclusions}
We have measured the binding energy of the $^{87}$Rb$^{133}$Cs molecule as
$h\times\unit{114\,268\,135\,237(5)(50)}{\kHz}$ using an optical frequency comb
based on difference-frequency
generation~\cite{Telle:2005,D.Fehrenbacher_Optica_2015}. The results for
different intermediate states $\unit{\sim1.65}{\THz}$ apart agree within their
experimental uncertainty and we are able to resolve the nuclear Zeeman splitting
of the molecular ground state. The accuracy of our ground-state binding energy
measurement is limited by uncertainties in the theoretical models of the
molecular structure. This is, to our knowledge, the most accurate determination
to date of the dissociation energy of a molecule. The ability to measure
molecular transitions with high precision is also potentially relevant to
searches for variations in fundamental constants such as the electron-proton
mass ratio~\cite{Demille2008}.
\nocite{Hughes2010}

\begin{acknowledgments}
We would like to thank the group of M.~P.~A.~Jones for providing the Sr $^1S_0
\rightarrow\,^3P_1$ reference for the wavemeter calibrations, I. G. Hughes for
useful discussions on error analysis and B.~Lu, Z.~Ji and M.~P.~K\"{o}ppinger
for their work on the early stages of the project. This work was supported by
the UK EPSRC [Grants EP/H003363/1, EP/I012044/1 and ER/S78339/01]. The
experimental data and analysis presented in this paper are available at INSERT
DOI ON PUBLICATION.
\end{acknowledgments}

\end{document}